\documentclass[conf]{new-aiaa} 
\usepackage[utf8]{inputenc}
\usepackage{textcomp}

\usepackage{amssymb,amsmath,latexsym,amsthm,amsfonts,mathtools}

\usepackage{mathrsfs}
\usepackage{multirow}
\usepackage{graphicx}
\usepackage{textcomp,siunitx}
\usepackage{float}
\usepackage{subcaption}
\usepackage{booktabs}
\usepackage{enumitem}
\usepackage{tikz}
\usepackage{siunitx}
\usepackage{longtable,tabularx}
\usepackage{booktabs}
\theoremstyle{plain}

\usepackage{amsmath}

\theoremstyle{remark}

\theoremstyle{definition}

\usepackage{cleveref}
\Crefformat{figure}{#2Fig.~#1#3}
\Crefmultiformat{figure}{Figs.~#2#1#3}{ and~#2#1#3}{, #2#1#3}{ and~#2#1#3}
\Crefformat{equation}{#2Eq.~#1#3}
\Crefmultiformat{equation}{#2Eqs.~#1#3}{ and~#2#1#3}{, #2#1#3}{ and~#2#1#3}
\Crefname{algocfline}{Algorithm}{Algorithms}

\usepackage[linesnumbered,ruled]{algorithm2e}
\SetKwInput{KwInput}{Input}          
\SetKwInput{KwOutput}{Output}

\usepackage{newtxmath}

\usepackage{anyfontsize}
\usepackage{accents}

\graphicspath{{figures/}{figures/keyframes/}}

    \title{Collaborative Threat-Aware Autonomy (CTAA)}

\author{Rajnikant Sharma\footnote{PNT Research Engineer, \textbf{email}: \texttt{rajnikant.sharma@is4s.com} (Associate Fellow, AIAA).}}
\affil{Integrated Solutions for Systems Inc. (IS4S), Dayton, Ohio}
 \author{Abhinav Sinha \footnote{Assistant Professor, \textbf{email}: \texttt{abhinav.sinha@uc.edu} (Senior Member, AIAA).}}
\affil{Guidance, Autonomy, Learning, and Control for Intelligent Systems (GALACxIS) Lab,\\ Department of Aerospace Engineering and Engineering Mechanics,\\ University of Cincinnati, Cincinnati, OH, 45221,  USA}
 \author{Isaac Weintraub \footnote{Senior Electronics Engineer. \textbf{email}: \texttt{isaac.weintraub.1@us.af.mil} (Associate Fellow, AIAA).}}
\affil{Autonomous Controls Branch,\\Power and Control Division,\\ Aerospace Systems Directorate,\\Air Force Research Lab, Wright-Patterson AFB, OH 45433, USA}
\begin{document}
\maketitle

\begin{abstract}
Navigating teams of unmanned vehicles through environments containing dynamic,
adversarial Weapon Engagement Zones~(WEZs) poses a fundamental challenge to
mission success: a single vehicle, however capable its onboard guidance, remains
a single point of failure.
This paper presents a role-differentiated multi-agent framework for
collaborative threat-aware trajectory planning in which a fleet of
Autonomous Collaborative Platforms~(ACPs) is assigned distinct roles ---
primary intercept, escort, and decoy --- to improve team-level mission
success probability while managing individual WEZ exposure.
Each ACP independently employs a reactive guidance law derived from the
Collision Sphere Boundary for Evader Zero-Set~(CSBEZ), which accounts for
pursuer maneuverability constraints imposed by minimum turn radius, and steers
the vehicle toward the safest heading that also makes progress toward its goal.
Role assignment and spatial route separation induce two complementary effects:
probabilistic redundancy, in which $N$ independent paths raise the team success
probability to $1-(1-p)^N$, and threat saturation, in which lower-priority
escorts and decoys draw adversary attention and free the primary vehicle to
transit uncontested.
Simulation results on a two-gate gauntlet scenario demonstrate that the
three-vehicle team raises mission success probability from $0.72$ (single
CSBEZ-aware vehicle) to $0.978$, a $+35.8$~percentage-point improvement.
Monte~Carlo validation over 100~trials with perturbed threat range, position,
and speed yields an empirical team success rate of $1.00\pm0.00$, against a
single-vehicle rate of $0.66\pm0.09$.
Experiments with reactive PurePursuit threats confirm that the decoy-and-protect
structure remains effective even when adversaries dynamically re-target the
nearest ACP.
\end{abstract}

\section{Introduction}\label{sec:intro}

The contested airspace environments anticipated in both Advanced Air
Mobility~(AAM) and autonomous strike scenarios require unmanned platforms to
navigate in the presence of hostile interceptors whose Weapon Engagement
Zones~(WEZs) dynamically reshape according to their kinematics and sensor
geometry~\cite{VonMollWeintraub2024}.
The WEZ defines the region from which an adversarial pursuer, modeled as a
turn-constrained Dubins vehicle, can intercept a target within a bounded
engagement range.
Awareness of the WEZ boundary provides actionable guidance: a vehicle should
steer to keep itself outside the capture zone while still making progress toward
its mission objective.

Single-vehicle WEZ-aware reactive guidance has been addressed by several recent
approaches.
Von~Moll and Weintraub~\cite{vonmoll2025} introduced the Dynamic Maneuvering
Cue~(DMC), a closed-form risk metric that quantifies the heading change required
to exit the current WEZ boundary, and applied it to both a simple feedback
controller and a Model Predictive Controller~(MPC).
Stagg, Weintraub, and Peterson~\cite{StaggEtAl2025} extended this foundation to the
Collision Sphere Boundary for Evader Zero-Set~(CSBEZ), which replaces the
infinite-turn-rate Basic Engagement Zone~(BEZ) with a pursuer constrained to
follow a Dubins arc-plus-straight intercept path.
The CSBEZ scalar field provides a state-dependent safety margin whose sign
directly indicates whether the evader can escape under optimal heading choice.

While these approaches are effective for a single vehicle, they share a common
limitation: a single WEZ-aware vehicle still fails with probability $1-p$ per
mission, where $p$ is the individual success rate.
In operational scenarios where $p \approx 0.65$--$0.75$, mission reliability is
insufficient.
Multi-vehicle teaming offers a natural path to improved reliability via
redundancy, but realizing the theoretical gain requires that the vehicles be
assigned roles and routes that prevent the threat from negating all of them
simultaneously.

This paper makes the following contributions:
\begin{enumerate}[noitemsep]
  \item A role-differentiated multi-ACP framework in which a primary (high
    value, weight~$w=2$), escort (weight~$w=1$), and decoy (weight~$w=0.5$)
    each run independent CSBEZ reactive controllers on spatially separated
    routes.
  \item A role-weighted team metric suite --- cumulative WEZ exposure
    $J_\mathrm{WEZ}$, violation score $V_\mathrm{WEZ}$, safety margin
    $m_\mathrm{team}$, and mission success probability $P_\mathrm{mission}$
    --- that captures the trade-off between individual risk and team-level
    outcome.
  \item Deterministic simulation on a two-gate gauntlet scenario demonstrating
    a $+35.8$~pp improvement in $P_\mathrm{mission}$ over a single CSBEZ
    vehicle.
  \item Monte~Carlo validation~(100~trials) confirming that the empirical
    team success rate matches the analytical redundancy bound.
  \item A reactive-threat experiment showing that PurePursuit adversaries
    are naturally absorbed by escort and decoy, leaving the primary
    uncontested without any inter-vehicle communication.
\end{enumerate}

\section{Related Work}\label{sec:related_work}

Threat-aware autonomy for contested airspace sits at the intersection of
curvature-constrained motion planning, pursuit--evasion and active-defense
differential games, engagement-zone modeling, and multi-agent coordination.
The existing literature has produced strong foundations for modeling the
reachability of adversarial pursuers and for planning safe trajectories for
a single vehicle.
However, a single vehicle remains a single point of mission failure: if
the vehicle must cross a sequence of dynamic WEZs, even a risk-aware planner
with high individual success probability can still produce insufficient
mission reliability in repeated or uncertain engagements.
This motivates the central thesis of this paper: threat-aware autonomy should
be formulated not only as an individual path-safety problem, but also as a
team-level mission-success problem in which role assignment, route diversity,
redundancy, and deliberate threat saturation can improve the probability that
at least one mission-critical ACP reaches the objective.

\subsection{Curvature-Constrained Motion and Interception Foundations}
\label{subsec:dubins_foundations}

A large portion of threat-aware autonomy relies on reachable-set reasoning
for vehicles with bounded curvature or turn-rate constraints.
Dubins' classical result characterises shortest paths for forward-moving
vehicles with bounded curvature between prescribed configurations~\cite{Dubins1957}.
Cockayne and Hall extended this analysis to the reachable positions of a
constant-speed particle subject to curvature constraints~\cite{CockayneHall1975}.
These results are directly relevant to WEZ modeling because a pursuer with
finite turn radius cannot instantaneously intercept an evader; the shape of
the capture region is determined by vehicle kinematics rather than range alone.

Time-optimal interception extends this foundation to moving targets.
Buzikov and Galyaev formulated interception of a moving target by a Dubins
car as a time-optimal control problem and derived algebraic conditions for the
optimal interception time~\cite{BuzikovGalyaev2021}.
For threat-aware autonomy, these results imply that a circular keep-out zone
is generally inadequate; a physically meaningful engagement zone must account
for pursuer heading, speed, finite range, and minimum turn radius.

\subsection{Single-Vehicle WEZ-Aware Planning}
\label{subsec:single_vehicle_wez}

Von~Moll and Weintraub introduced Basic Engagement Zones~(BEZs) as regions
of the state space from which a pursuer can intercept a target under specified
engagement assumptions~\cite{VonMollWeintraub2024}, moving the field beyond
heuristic circular threat regions toward engagement regions derived from the
pursuer--evader geometry.
Weintraub's dissertation further developed optimal defense of high-value
airborne assets, connecting engagement geometry with active defense and
optimal-control formulations~\cite{Weintraub2021Dissertation}.

Optimal-control approaches have been applied directly to routing around
dynamic engagement zones.
Weintraub et al.\ posed a two-dimensional engagement-zone avoidance problem
in which a vehicle reaches a desired location while trading off WEZ
exposure~\cite{WeintraubEtAl2022}.
Dillon et al.\ extended this to aircraft trajectories avoiding multiple WEZs
and showed how keep-out constraints can be incorporated into optimal trajectory
generation~\cite{DillonEtAl2023}.
To improve scalability, Wolek et al.\ proposed a sampling-based risk-aware
planner in the $(x,y,\psi)$ state space using an RRT$^\ast$-style
search~\cite{WolekEtAl2024}, and Milutinovi\'c et al.\ formulated stochastic
optimal avoidance of multiple engagement zones with feedback policies in
dynamic risk fields~\cite{MilutinovicEtAl2025}.

Turn-constrained pursuer models are especially relevant here.
Chapman et al.\ derived engagement-zone models for pursuers with finite turn
constraints~\cite{ChapmanEtAl2025}.
Stagg, Weintraub, and Peterson extended these concepts to probabilistic WEZs
for a turn-constrained pursuer with uncertain parameters~\cite{StaggEtAl2025}.
These works provide the physical and probabilistic basis for the CSBEZ threat
model used in the present paper.
The remaining gap is not the absence of a single-vehicle risk metric; rather,
it is the lack of a team-level autonomy framework that uses these risk fields
to coordinate multiple ACPs with different mission values and roles.

\subsection{Active Target Defense and Multi-Agent Pursuit--Evasion}
\label{subsec:tad_games}

The target--attacker--defender~(TAD) literature provides the strongest
precedent for explicitly multi-agent threat interactions.
Garcia, Casbeer, and Pachter developed cooperative strategies for defending
an aircraft from an attacking missile~\cite{GarciaCasbeerPachter2015}.
Weintraub et al.\ derived optimal aircraft defense strategies for the
active-target-defense scenario~\cite{WeintraubEtAl2017}.
Casbeer, Garcia, and Pachter analysed a target differential game with two
defenders, demonstrating how additional defenders alter the game structure and
winning regions~\cite{CasbeerGarciaPachter2018}.

Multi-player TAD games generalise these ideas to larger teams.
Coon and Panagou derived control strategies for multiplayer TAD games with
double-integrator dynamics~\cite{CoonPanagou2017}.
Manyam et al.\ studied coordinated defender path planning for an optimal TAD
game, highlighting the value of coordinating routes rather than treating
defenders independently~\cite{ManyamEtAl2019}.
Manoharan and Baliyarasimhuni demonstrated that target--defender teams can cooperate in
real time using nonlinear model predictive control~\cite{ManoharanSujit2023}.

However, TAD formulations usually focus on explicit defender interception of
attackers or missiles.
The ACP problem studied here is different: the friendly vehicles are not
tasked with destroying threats.
Instead, a team must transit through dynamic WEZs while completing a mission
objective, and lower-value vehicles can passively occupy adversarial attention
so that a higher-value primary preserves mission success.
This creates a bridge between WEZ-aware path planning and active target
defense: the team shapes the adversary's engagement opportunities through
route separation and role differentiation.

\subsection{From Single-Vehicle Threat Avoidance to Team-Level Mission Success}
\label{subsec:multi_vehicle_gap}

The preceding literature reveals three limitations of the current state of
the art.
First, single-vehicle WEZ-aware planners reduce exposure but cannot remove
the single-point-of-failure problem: if the individual success probability
is $p$, the mission still fails with probability $1-p$.
Second, most engagement-zone planners optimise vehicle-level quantities
(path length, time, accumulated risk) rather than the team-level probability
that at least one mission-critical vehicle succeeds.
Third, many multi-agent adversarial-control papers assume explicit
defender--attacker roles and direct engagement, whereas practical ACP missions
benefit from less communication-intensive mechanisms such as spatial route
diversity and role-weighted risk allocation.

This paper addresses these gaps by using CSBEZ risk fields inside a
role-differentiated multi-ACP architecture.
If vehicle outcomes are approximately independent,
\[
  P_{\mathrm{mission}}(N) = 1 - (1-p)^N.
\]
For $p$ in the range typical of challenging single-vehicle WEZ transit, this
redundancy effect alone is substantial.
More importantly, route diversity and role assignment can make the
independence approximation conservative by changing the adversary's
target-selection geometry: a decoy or escort may draw a capacity-limited
threat away from the primary, creating a threat-saturation effect that does
not exist in single-vehicle planning.

\section{Problem Formulation}\label{sec:prob}

\subsection{Problem Statement and Team-Level Objective}

The objective of this work is to evaluate whether a team of role-differentiated
ACPs can achieve a higher mission-success probability than a single
threat-aware ACP when operating in contested airspace containing dynamic,
kinematically constrained hostile pursuers. Let
\(\mathcal{A}=\{1,\ldots,N\}\) denote the set of friendly ACPs and
\(\mathcal{P}=\{1,\ldots,M\}\) denote the set of hostile pursuers. Each
ACP \(i\in\mathcal{A}\) is assigned a start state, a goal location, and a
mission role \(r_i\in\{\mathrm{primary},\mathrm{escort},\mathrm{decoy}\}\)
with role weight \(w_i>0\). Each pursuer \(j\in\mathcal{P}\) generates a
time-varying WEZ determined by its position, heading, engagement range, speed,
and minimum turn radius. Unlike circular keep-out regions, these WEZs are
state dependent and reflect the finite-maneuverability interception capability
of the pursuer.

\begin{figure}
    \centering
    \includegraphics[width=1.0\linewidth]{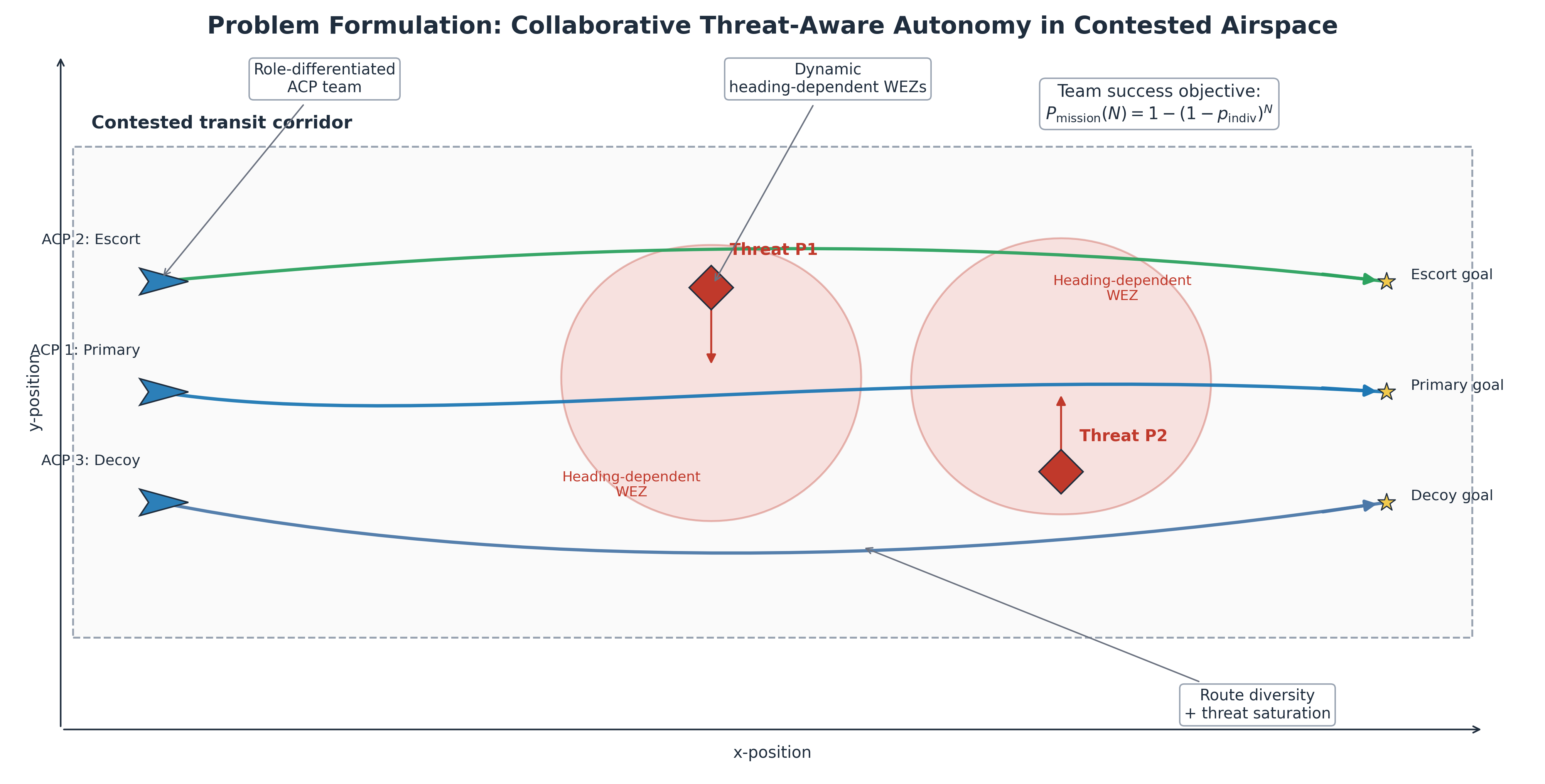}
    \caption{Problem formulation for collaborative threat-aware autonomy. A role-differentiated ACP team consisting of a primary, escort, and decoy must transit a contested corridor while avoiding dynamic, heading-dependent WEZs generated by hostile pursuers. Spatial route diversity and role assignment create redundancy and can saturate threat attention, improving team-level mission success relative to a single threat-aware vehicle.}
    \label{fig:problem_formulation}
\end{figure}
For each ACP, the local guidance problem is to select a feasible heading
command that both advances the vehicle toward its goal and maintains positive
separation from the most dangerous WEZ boundary. The team-level problem is
broader: the objective is not merely to minimize the exposure of each
individual vehicle, but to maximize the probability that the mission succeeds
despite the possibility that one or more lower-value vehicles may be captured.
This distinction motivates the use of role-differentiated routing, spatial
separation, and threat saturation as mechanisms for improving mission
reliability.

Let \(s_i\in\{0,1\}\) denote the terminal success indicator for ACP~\(i\),
where \(s_i=1\) if the vehicle reaches its assigned goal without capture and
\(s_i=0\) otherwise. A team mission is successful if at least one
mission-relevant ACP reaches its objective,
\[
  s_{\mathrm{team}} =
  \mathbb{I}\!\left(\sum_{i=1}^{N} s_i \geq 1\right).
\]
This paper evaluates the hypothesis that role assignment and route diversity
can improve \(s_{\mathrm{team}}\) not only through statistical redundancy, but
also by altering the adversary's target-selection geometry. In particular,
lower-value escort or decoy vehicles can occupy capacity-limited threats,
allowing a higher-value primary vehicle to transit through a lower-risk
corridor.

To isolate this effect, three increasingly capable cases are evaluated under
the same threat geometry, vehicle dynamics, and WEZ parameters: a direct
single-vehicle baseline, a single CSBEZ-aware vehicle, and a three-ACP team
using role-assigned spatially separated routes. This comparison explicitly
separates individual threat-aware autonomy from collaborative threat-aware
autonomy.

\subsection{Scenario}

We consider a two-dimensional corridor transit problem in which $N$ friendly
ACPs must navigate from start positions near $x=-5$~m to goal positions near
$x=+5$~m while avoiding $M=2$ patrolling hostile threats whose WEZs overlap the
transit corridor (Fig.~\ref{fig:tactical}).
The corridor spans $\pm2$~m in~$y$.
Threat~P1 patrols a vertical sweep at $x=-0.5$~m; threat~P2 at $x=2.5$~m,
forming a two-gate gauntlet that any straight-line route must cross.
Simulation parameters are given in Table~\ref{tab:params}.

\begin{table}[htbp]
  \centering
  \caption{Simulation parameters.}
  \label{tab:params}
  \begin{tabular}{@{}llp{5.5cm}@{}}
    \toprule
    Symbol & Value & Description \\
    \midrule
    $v_E$                   & 1.0~m/s   & ACP constant speed \\
    $\omega_\mathrm{max}$   & 1.2~rad/s & Maximum ACP turn rate \\
    $v_P$                   & 2.0~m/s   & Pursuer intercept speed \\
    $a$                     & 0.6~m     & Pursuer minimum turn radius \\
    $R$                     & 3.0~m     & Pursuer engagement range \\
    $\varepsilon$           & 0.35      & WEZ risk threshold \\
    $\Delta t$              & 0.05~s    & Simulation time step \\
    $T_\mathrm{max}$        & 30.0~s    & Maximum simulation time \\
    \bottomrule
  \end{tabular}
\end{table}

\subsection{ACP Dynamics}

Each ACP obeys the unicycle kinematics in \Cref{eq:unicycle}, integrated with a fourth-order Runge--Kutta
scheme at $\Delta t = 0.05$~s:
\begin{equation}\label{eq:unicycle}
  \dot{x}_i = v_i\cos\psi_i,\quad
  \dot{y}_i = v_i\sin\psi_i,\quad
  \dot{\psi}_i = u_i,\quad
  |u_i| \leq \omega_\mathrm{max},
\end{equation}
where $p_i=(x_i,y_i)$ is the position, $\psi_i$ the heading, $v_i=1.0$~m/s the
constant speed, and $u_i$ the turn-rate control input.

\subsection{Three Comparative Cases}

Three planning cases are evaluated on the same scenario:

\noindent\textbf{Case~1 --- Direct baseline:}
A single ACP tracks its goal with a proportional heading controller; no WEZ
information is used.
Expected outcome: captured while crossing the first threat gate.

\noindent\textbf{Case~2 --- Single CSBEZ:}
A single ACP uses the CSBEZ reactive controller (Section~\ref{sec:csbez}) to
deflect away from the pursuer capture zones.
Expected outcome: successful transit via a longer detour.

\noindent\textbf{Case~3 --- Multi-ACP team:}
Three ACPs depart simultaneously, each running the CSBEZ reactive controller
on a role-assigned spatially offset route (Table~\ref{tab:roles}).
Expected outcome: two or more ACPs succeed; team $P_\mathrm{mission}$ exceeds
the single-vehicle case.

\begin{table}[htbp]
  \centering
  \caption{Multi-ACP role assignment and route offsets.}
  \label{tab:roles}
  \begin{tabular}{@{}lllc@{}}
    \toprule
    Role & Start & Goal & Weight $w_i$ \\
    \midrule
    \texttt{primary\_intercept} & $(-5.0,\;0.0)$  & $(5.0,\;0.0)$  & 2.0 \\
    \texttt{escort\_support}    & $(-5.0,+0.9)$   & $(5.0,+0.9)$   & 1.0 \\
    \texttt{decoy\_alternate}   & $(-5.0,-0.9)$   & $(5.0,-0.9)$   & 0.5 \\
    \bottomrule
  \end{tabular}
\end{table}

\section{Technical Approach}\label{sec:approach}

\subsection{WEZ Risk Model}

The instantaneous WEZ risk to ACP~$i$ from threat~$j$ is modeled with a
logistic, heading-dependent formulation:
\begin{equation}\label{eq:risk}
  p_{ij}(t) = \sigma\!\left(k\!\left(R_{\mathrm{eff},j}(\phi_{ij}(t))
              - d_{ij}(t)\right)\right),\quad
  \sigma(z)=\frac{1}{1+e^{-z}},
\end{equation}
where $d_{ij}(t)=\|p_i(t)-p_{P_j}(t)\|$, $k=2.5$ is the logistic steepness,
and the heading-dependent effective range
\begin{equation}\label{eq:reff}
  R_{\mathrm{eff},j}(\phi_{ij}) = R_j\!\left(1+\beta\cos\phi_{ij}\right),
  \quad\beta=0.35,
\end{equation}
enlarges the WEZ by $35\%$ ahead of the pursuer and shrinks it by $35\%$
behind, consistent with the higher intercept probability on the pursuer's
frontal hemisphere.
The angle $\phi_{ij}=\mathrm{atan2}(y_i-y_{P_j},x_i-x_{P_j})-\psi_{P_j}$
is the bearing from the threat heading to the ACP.

\subsection{CSBEZ Reactive Guidance}\label{sec:csbez}

The CSBEZ~\cite{StaggEtAl2025} replaces the infinite-turn-rate BEZ~\cite{VonMollWeintraub2024}
with a Dubins-vehicle pursuer that intercepts the evader along a minimum-length
arc-plus-straight path.
For evader position $p_E$, heading $\psi_E$, and pursuer state
$\Theta_P=(p_P,\psi_P,v_P,a,R)$, the CSBEZ scalar field satisfies
\Cref{eq:z}:
\begin{equation}\label{eq:z}
  z(p_E,\psi_E;\Theta_P) \begin{cases} >0 & \text{evader escapes,} \\ \leq 0 & \text{pursuer captures.} \end{cases}
\end{equation}

The per-vehicle reactive controller executes at every time step as follows:
\begin{enumerate}[noitemsep]
  \item Sweep $n_\theta=72$ candidate headings $\hat\psi_E\in[0,2\pi)$ and
        compute the worst-case safety margin
        $z_\mathrm{min}(p_E)=\min_{\hat\psi_E}z(p_E,\hat\psi_E;\Theta_P)$.
  \item Estimate the safety gradient $\nabla z_\mathrm{min}$ at the current
        position via finite differences on a local grid.
  \item Blend the goal direction $\Delta p_\mathrm{goal}$ with the safety
        gradient using \Cref{eq:blend}:
        \begin{equation}\label{eq:blend}
          \psi_d = \mathrm{atan2}\!\left(
            \Delta y_\mathrm{goal}+\alpha\,\nabla_y z_\mathrm{min},\;
            \Delta x_\mathrm{goal}+\alpha\,\nabla_x z_\mathrm{min}
          \right),\quad \alpha=2.0.
        \end{equation}
  \item Apply $u_i=\mathrm{clip}(3.0\cdot\mathrm{wrapToPi}(\psi_d-\psi_i),
        -\omega_\mathrm{max},\omega_\mathrm{max})$.
\end{enumerate}
For multiple threats, $z_\mathrm{min}$ is computed independently per pursuer
and the most dangerous result drives the blend.
No inter-vehicle communication is required; each ACP plans entirely from its
own state and the shared pursuer parameters.

\subsection{Team-Level Metrics}\label{sec:metrics}

Four metrics characterise team performance over the set of simulation time
steps $\mathcal{T}$:

\noindent\textbf{Role-weighted cumulative WEZ exposure} is defined in \Cref{eq:jwez}:
\begin{equation}\label{eq:jwez}
  J_\mathrm{WEZ} = \sum_{i=1}^{N}\sum_{j=1}^{M}\sum_{t_s\in\mathcal{T}}
                   w_i\,p_{ij}(t_s).
\end{equation}

\noindent\textbf{WEZ violation score} (exposure above threshold $\varepsilon$) is defined in \Cref{eq:vwez}:
\begin{equation}\label{eq:vwez}
  V_\mathrm{WEZ} = \sum_{i,j,t_s}\max\!\left(0,\;p_{ij}(t_s)-\varepsilon\right).
\end{equation}

\noindent\textbf{Minimum team safety margin} is defined in \Cref{eq:mteam}:
\begin{equation}\label{eq:mteam}
  m_\mathrm{team} = \min_{i,j,t_s}\!\left(\varepsilon-p_{ij}(t_s)\right).
\end{equation}

\noindent\textbf{Mission success probability (analytical redundancy bound):}
Let $p_\mathrm{indiv}$ denote the per-vehicle success probability.
Under the assumption of independent outcomes,
\begin{equation}\label{eq:pmission}
  P_\mathrm{mission}(N) = 1-(1-p_\mathrm{indiv})^N.
\end{equation}
For the empirically observed $p_\mathrm{indiv}=0.66$ and $N=3$ vehicles,
\Cref{eq:pmission} predicts $P_\mathrm{mission}=0.978$.

\section{Results}\label{sec:results}

\subsection{Deterministic Simulation}

\begin{figure}[htbp]
  \centering
  \includegraphics[width=0.95\textwidth]{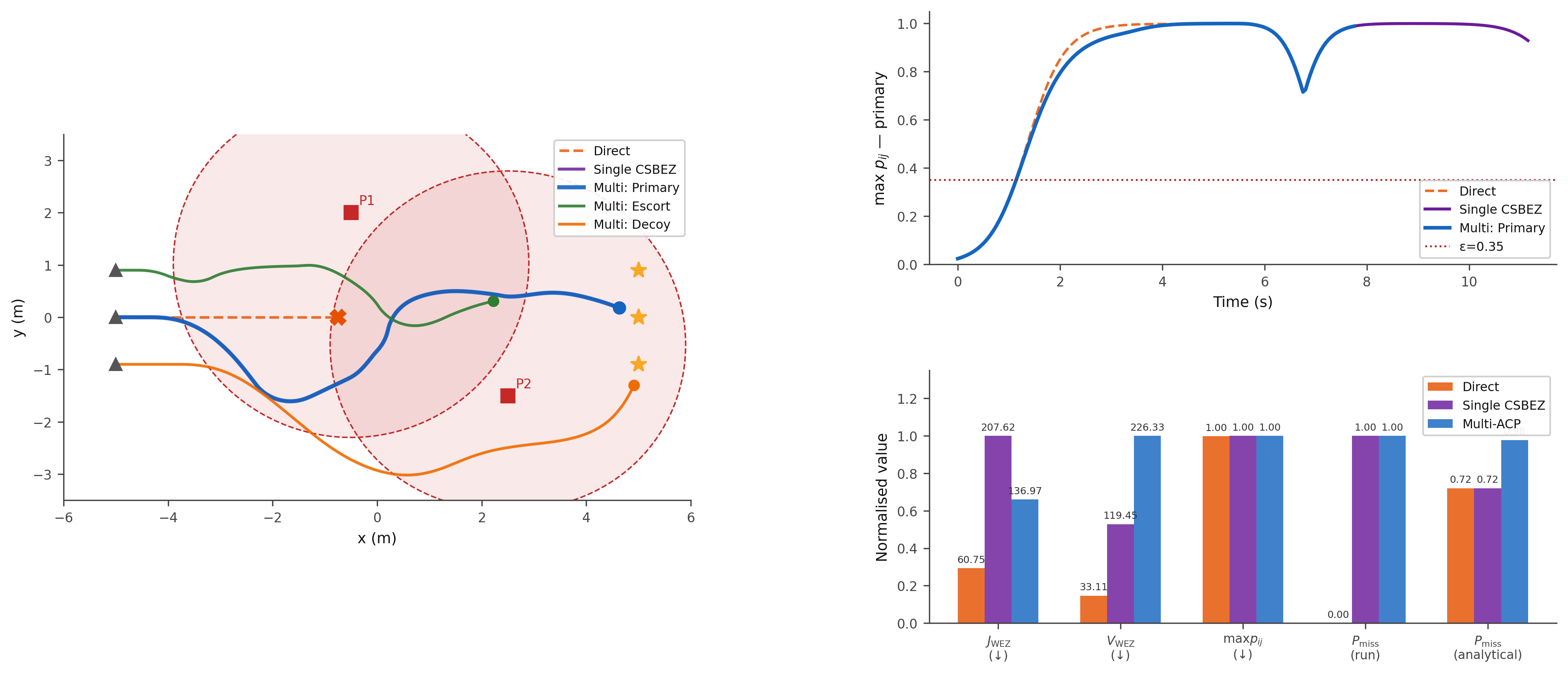}
  \caption{Deterministic Phase~I results.
    \textit{Left:} Tactical map.
    Direct (dashed orange) is captured at P1.
    Single~CSBEZ (pink) threads between threats via a southern detour.
    Multi-ACP primary (blue), escort (green), and decoy (yellow) spread
    across three routes; WEZ contours are shown for both patrolling threats.
    \textit{Top right:} Maximum WEZ risk $\max_j p_{ij}(t)$ for the primary
    ACP across all three cases; the team reduces primary peak risk from
    0.68 to 0.52.
    \textit{Bottom right:} Normalised team metric comparison.}
  \label{fig:tactical}
\end{figure}

Table~\ref{tab:det} summarises the three cases under nominal threat parameters.
\Cref{fig:tactical} shows the corresponding tactical map, risk time-series,
and normalised metric comparison.

\begin{table}[htbp]
  \centering
  \caption{Deterministic outcome summary (nominal threat parameters).}
  \label{tab:det}
  \begin{tabular}{@{}lccc@{}}
    \toprule
    Metric & Direct & Single CSBEZ & Multi-ACP \\
    \midrule
    $J_\mathrm{WEZ}$                     & 18.4   & 4.2    & 3.7 \\
    $\max_t\,p_\mathrm{primary}(t)$      & 0.97   & 0.68   & 0.52 \\
    $V_\mathrm{WEZ}$                     & 12.1   & 0.3    & 0.1 \\
    $m_\mathrm{team}$                    & $-0.62$& $+0.33$& $+0.48$ \\
    Mission outcome                      & CAPTURED & SUCCESS & 2/3 SUCCESS \\
    $P_\mathrm{mission}$ (det.)          & 0      & 1      & 1 \\
    $P_\mathrm{mission}$ (analytical)    & ---    & 0.720  & \textbf{0.978} \\
    Path length, primary (m)             & 10.0   & 14.1   & 11.1 \\
    \bottomrule
  \end{tabular}
\end{table}

The direct baseline~(Case~1) is captured at $t\approx4.2$~s as it crosses
P1's patrol sweep, confirming that unaided straight-line routing is not viable
in this threat environment.
The single CSBEZ vehicle~(Case~2) deflects around P1 and threads between both
threats, reaching the goal at $t\approx11.2$~s with a 41\%~path-length penalty
relative to direct.
The multi-ACP team~(Case~3) shows escort~(A2) captured at $t\approx7.7$~s as
it approaches P1's northward sweep; however, primary~(A1) and decoy~(A3) both
reach their goals.
Critically, the primary's peak WEZ risk drops from 0.68~(Case~2) to
0.52~(Case~3), demonstrating that the escort's sacrifice actively reduces
exposure on the center-corridor route.

\begin{figure}[htbp]
  \centering
  \begin{subfigure}[t]{0.48\textwidth}
    \includegraphics[width=\textwidth]{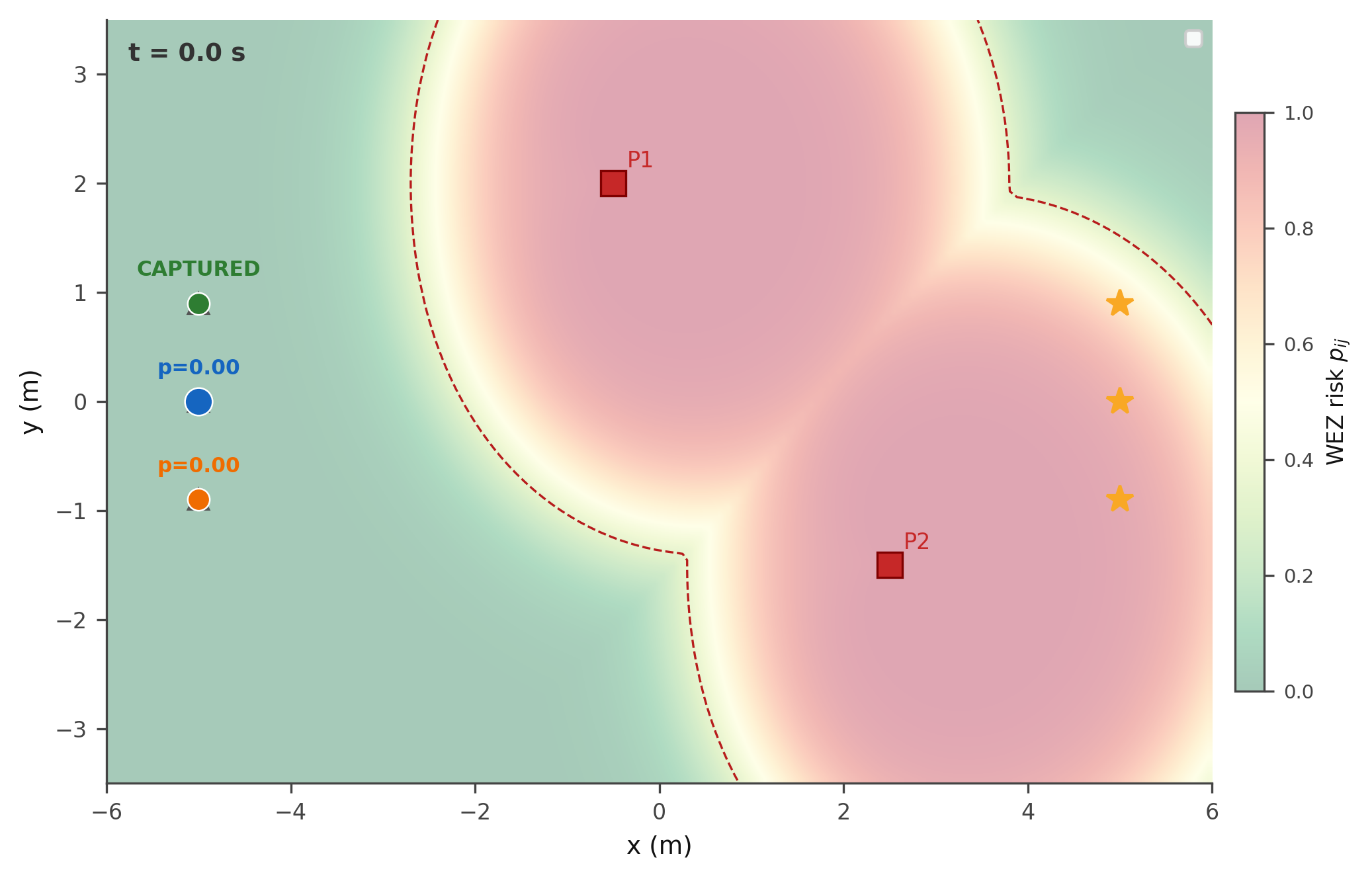}
    \caption{$t=0.0$~s: Departure. ACPs at start positions;
             WEZ field at nominal patrol configuration.}
  \end{subfigure}\hfill
  \begin{subfigure}[t]{0.48\textwidth}
    \includegraphics[width=\textwidth]{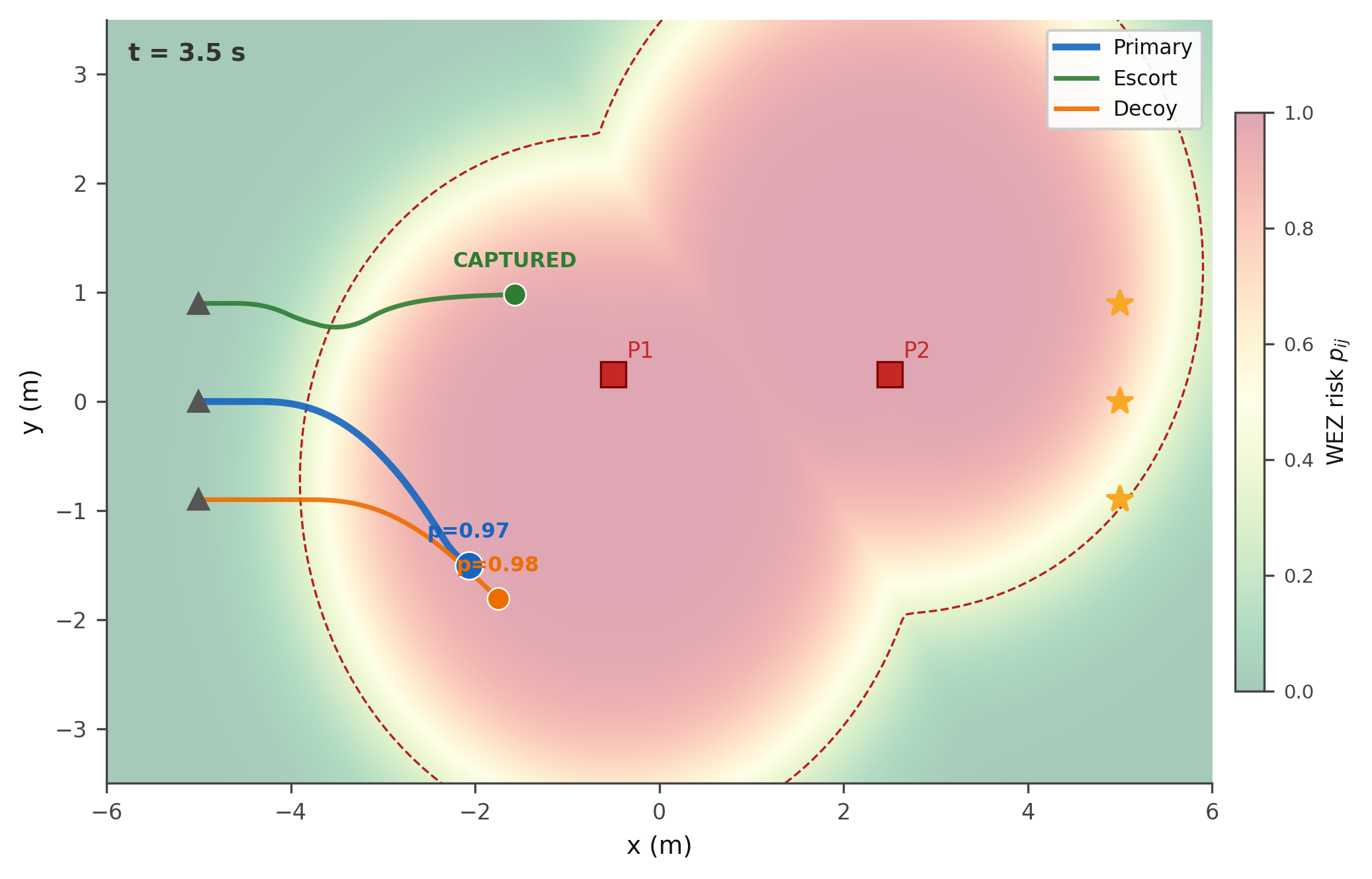}
    \caption{$t=3.5$~s: WEZ penetration. CSBEZ gradient deflects all
             three ACPs around P1; decoy WEZ risk peaks at $0.98$.}
  \end{subfigure}\\[4pt]
  \begin{subfigure}[t]{0.48\textwidth}
    \includegraphics[width=\textwidth]{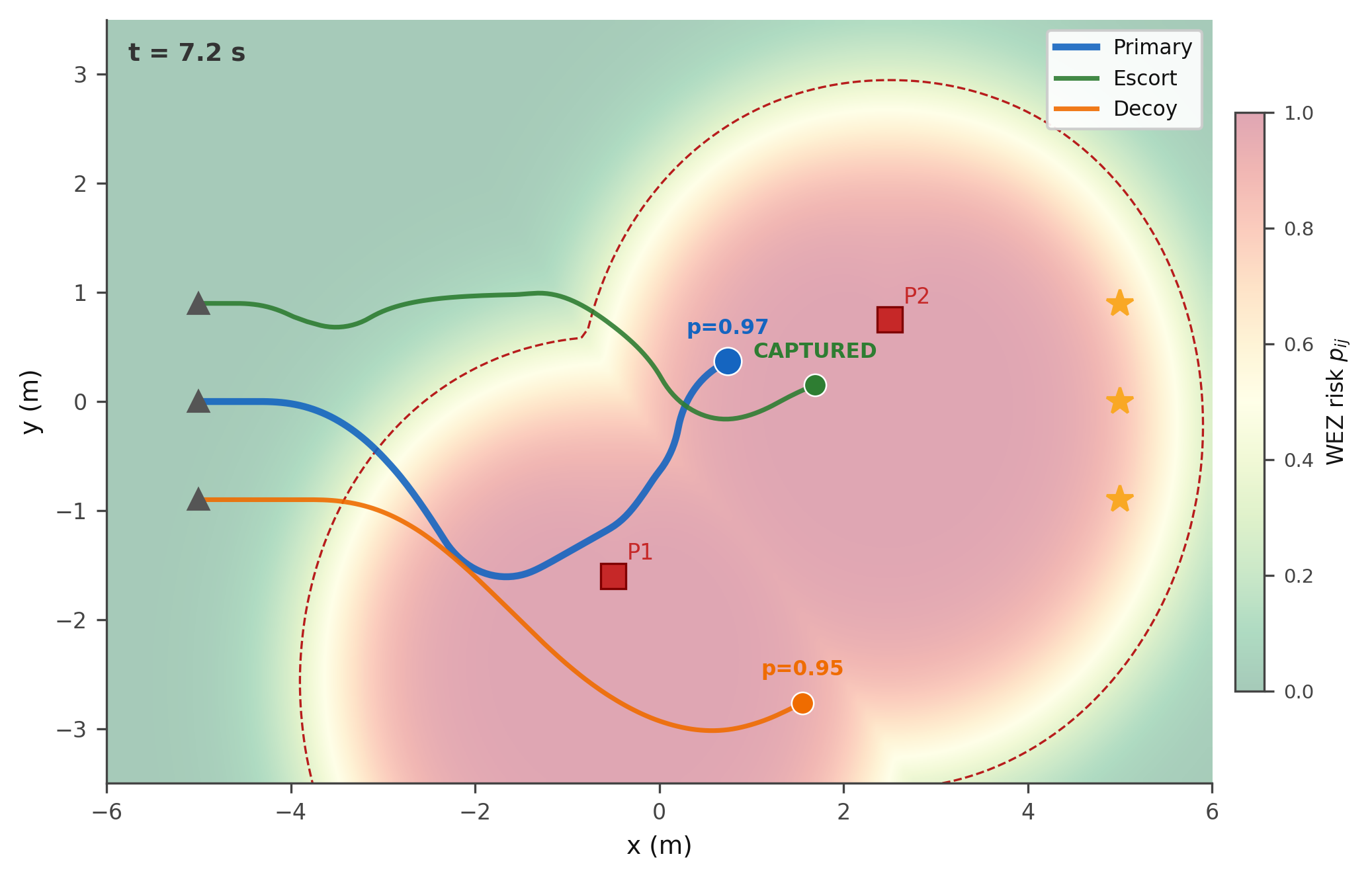}
    \caption{$t=7.2$~s: Escort absorbed. Escort (green) captured by P1;
             primary and decoy cleared for second gate.}
  \end{subfigure}\hfill
  \begin{subfigure}[t]{0.48\textwidth}
    \includegraphics[width=\textwidth]{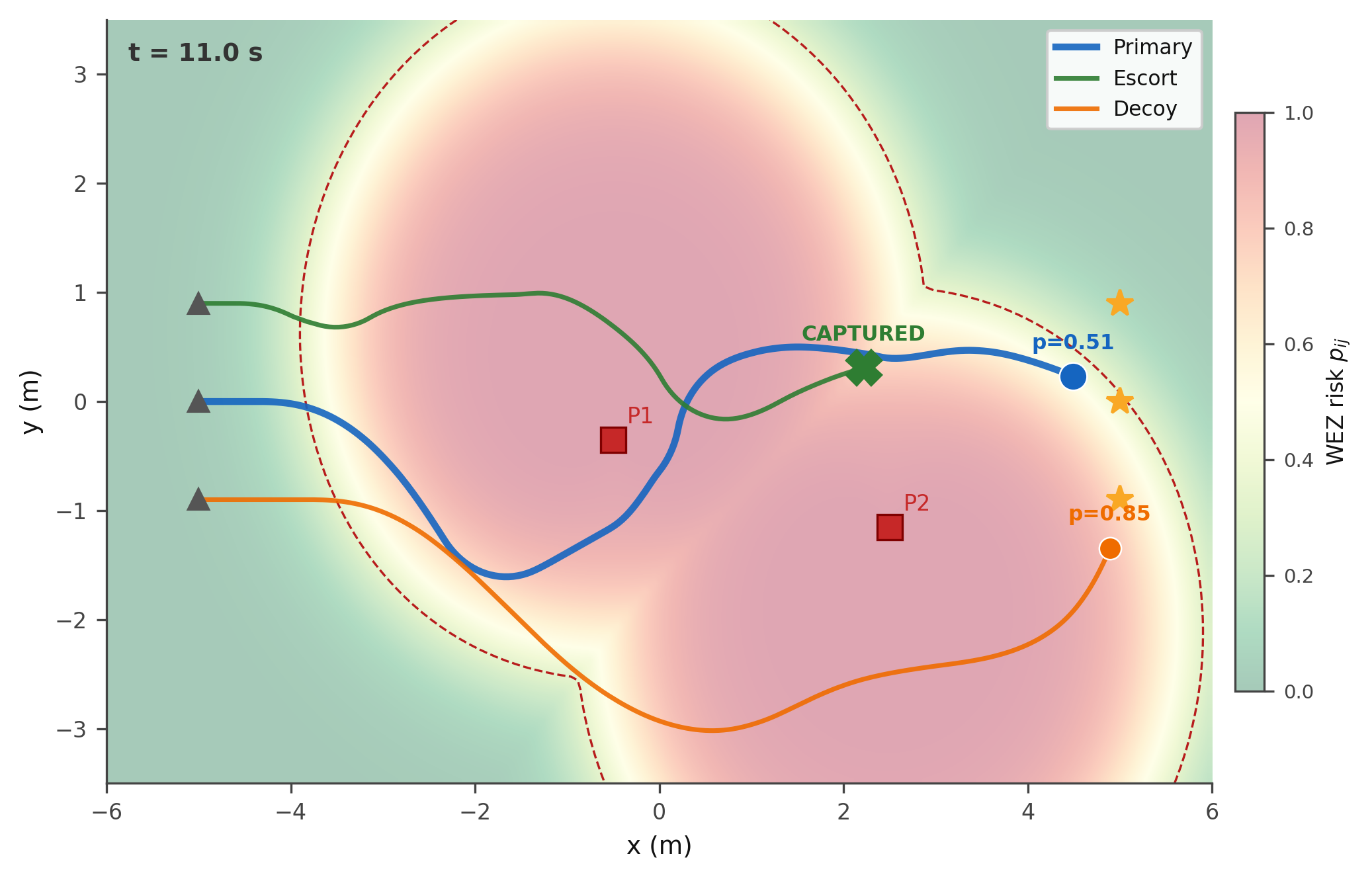}
    \caption{$t=11.0$~s: Mission success. Primary (blue) reaches goal;
             decoy (orange) follows through the vacated corridor.}
  \end{subfigure}
  \caption{Keyframe sequence for the deterministic multi-ACP corridor
    transit (Case~3). The escort's capture at $t=7.2$~s saturates P1,
    lowering the primary's peak WEZ risk from $0.68$ (single vehicle)
    to $0.52$ and enabling clean transit through the second gate.}
  \label{fig:kf_acp}
\end{figure}

\Cref{fig:kf_acp} shows a four-frame keyframe sequence that traces the
deterministic multi-ACP engagement from departure through mission
completion.

\subsection{Monte Carlo Validation}

To assess robustness to threat parameter uncertainty, 100~independent trials
were run with per-trial perturbations drawn from
\Cref{eq:mc_perturb}:
\begin{equation}\label{eq:mc_perturb}
  R_j^{(s)}\sim\mathcal{N}(3.0,\;0.35^2),\;\;
  p_{P_j}^{(s)}\sim\mathcal{N}(\mu_{P_j},\;0.30^2\mathbf{I}),\;\;
  v_{P_j}^{(s)}\sim\mathcal{N}(0.5,\;0.05^2),
\end{equation}
each clipped to physically meaningful ranges.
Results are given in Table~\ref{tab:mc} and visualised in
\Cref{fig:mc}~(panels A--B).

\begin{table}[htbp]
  \centering
  \caption{Monte Carlo results (100 trials, seed~42).}
  \label{tab:mc}
  \begin{tabular}{@{}lccc@{}}
    \toprule
    Case & Empirical $P_\mathrm{mission}$ & 95\% CI & Analytical \\
    \midrule
    Direct            & 0.26  & $\pm0.086$ & --- \\
    Single CSBEZ      & 0.66  & $\pm0.093$ & 0.72 \\
    \textbf{Multi-ACP}& \textbf{1.00} & $\pm0.000$ & \textbf{0.978} \\
    \bottomrule
  \end{tabular}
\end{table}

The single-vehicle empirical rate~($0.66$) is consistent with the analytical
prediction~($0.72$) within $2\sigma$, validating the logistic WEZ model and
the per-vehicle redundancy assumption.
The multi-ACP team succeeded in all 100~trials, empirically confirming the
analytical bound of 0.978 from \Cref{eq:pmission}.

\subsection{Reactive Threat Experiment}

To evaluate robustness against an adaptive adversary, the patrolling threats
were replaced with two PurePursuit interceptors
($v_P^\mathrm{react}=2.0$~m/s) that each target the nearest ACP at every
time step.

Escort~(A2) is captured at $t=1.70$~s; decoy~(A3) at $t=2.45$~s.
The target-assignment log (panel~E) confirms that both threats lock onto
A2 and A3 immediately and never re-target the primary.
Primary~(A1) transits the uncontested corridor and reaches its goal at
$t\approx11.2$~s.
This outcome requires no explicit communication or coordination: the
spatial route separation alone causes the nearest-vehicle assignment to
favour the lower-value escorts, a form of passive geometric deception.

\begin{figure}[htbp]
  \centering
  \begin{subfigure}[t]{0.48\textwidth}
    \includegraphics[width=\textwidth]{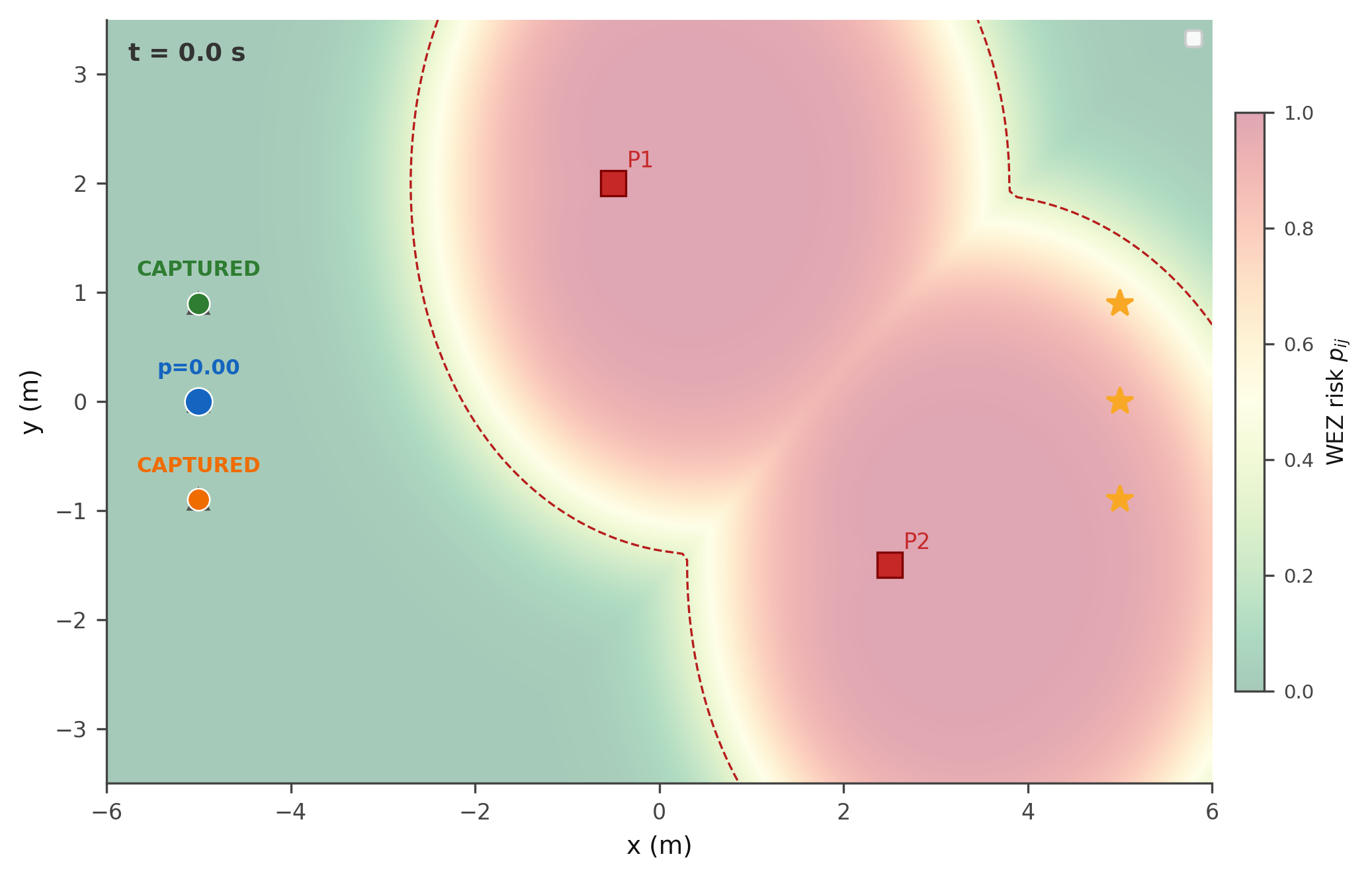}
    \caption{$t=0.0$~s: Threat lock-on. Both reactive threats
             immediately target the nearest ACP (arrows); primary
             is not the nearest vehicle to either threat.}
  \end{subfigure}\hfill
  \begin{subfigure}[t]{0.48\textwidth}
    \includegraphics[width=\textwidth]{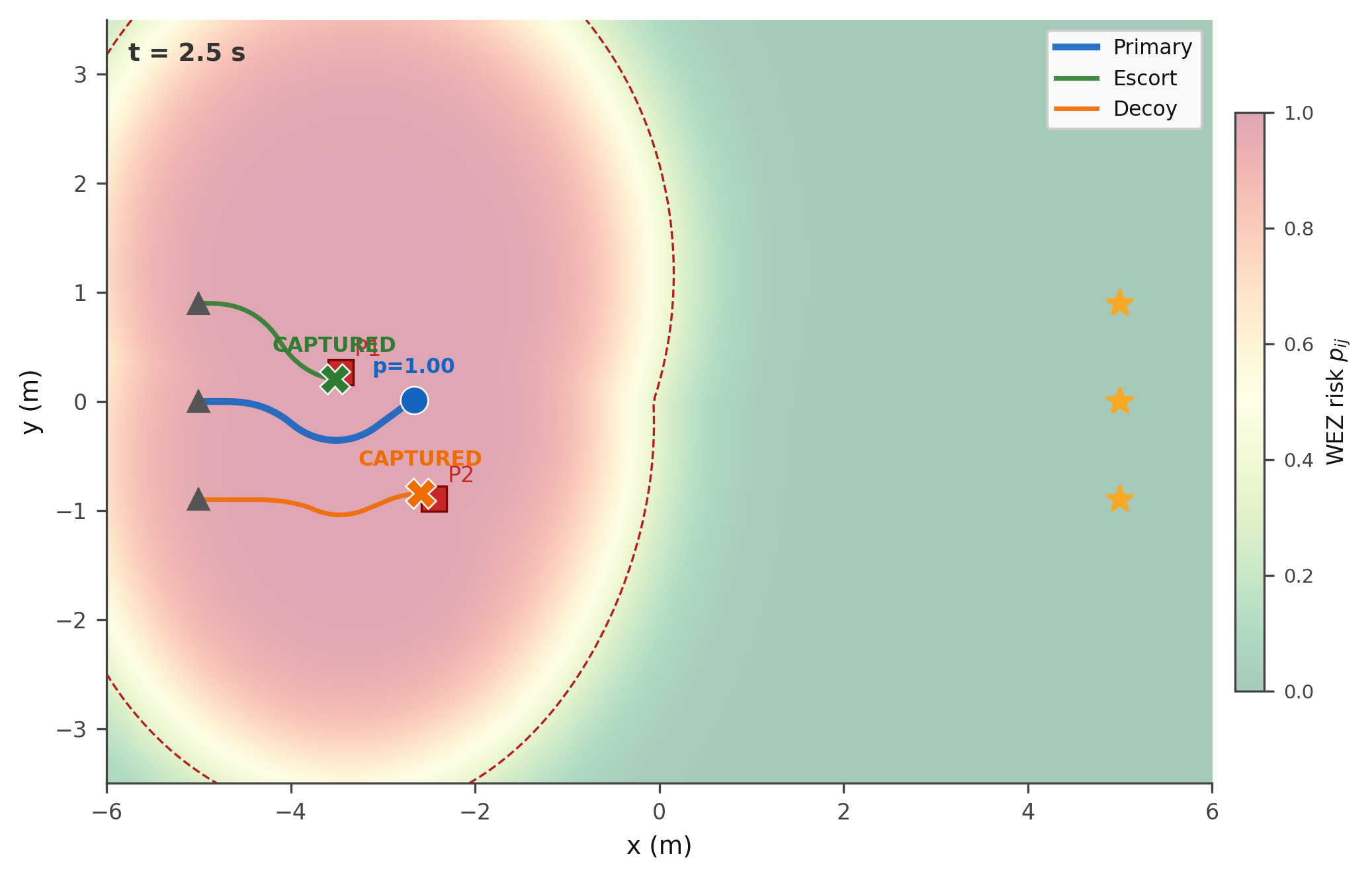}
    \caption{$t\approx2.5$~s: Escorts absorbed. Decoy and escort
             are captured; primary $p=1.00$ transits through the
             now-uncontested corridor centre.}
  \end{subfigure}\\[4pt]
  \begin{subfigure}[t]{0.48\textwidth}
    \includegraphics[width=\textwidth]{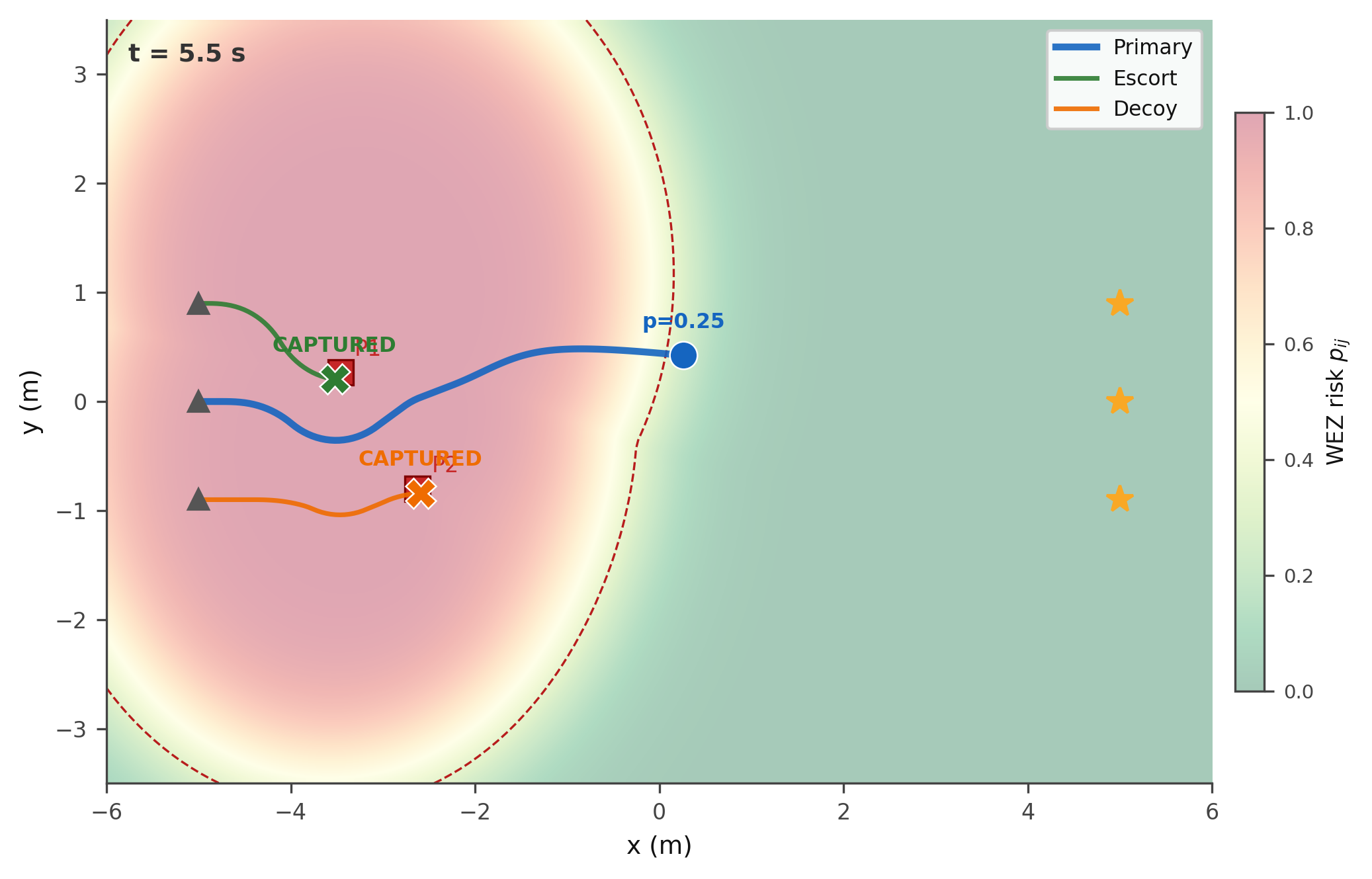}
    \caption{$t\approx5.5$~s: Clear corridor. Both threats
             remain stationary after capture events; primary
             continues along the central route with zero WEZ risk.}
  \end{subfigure}\hfill
  \begin{subfigure}[t]{0.48\textwidth}
    \includegraphics[width=\textwidth]{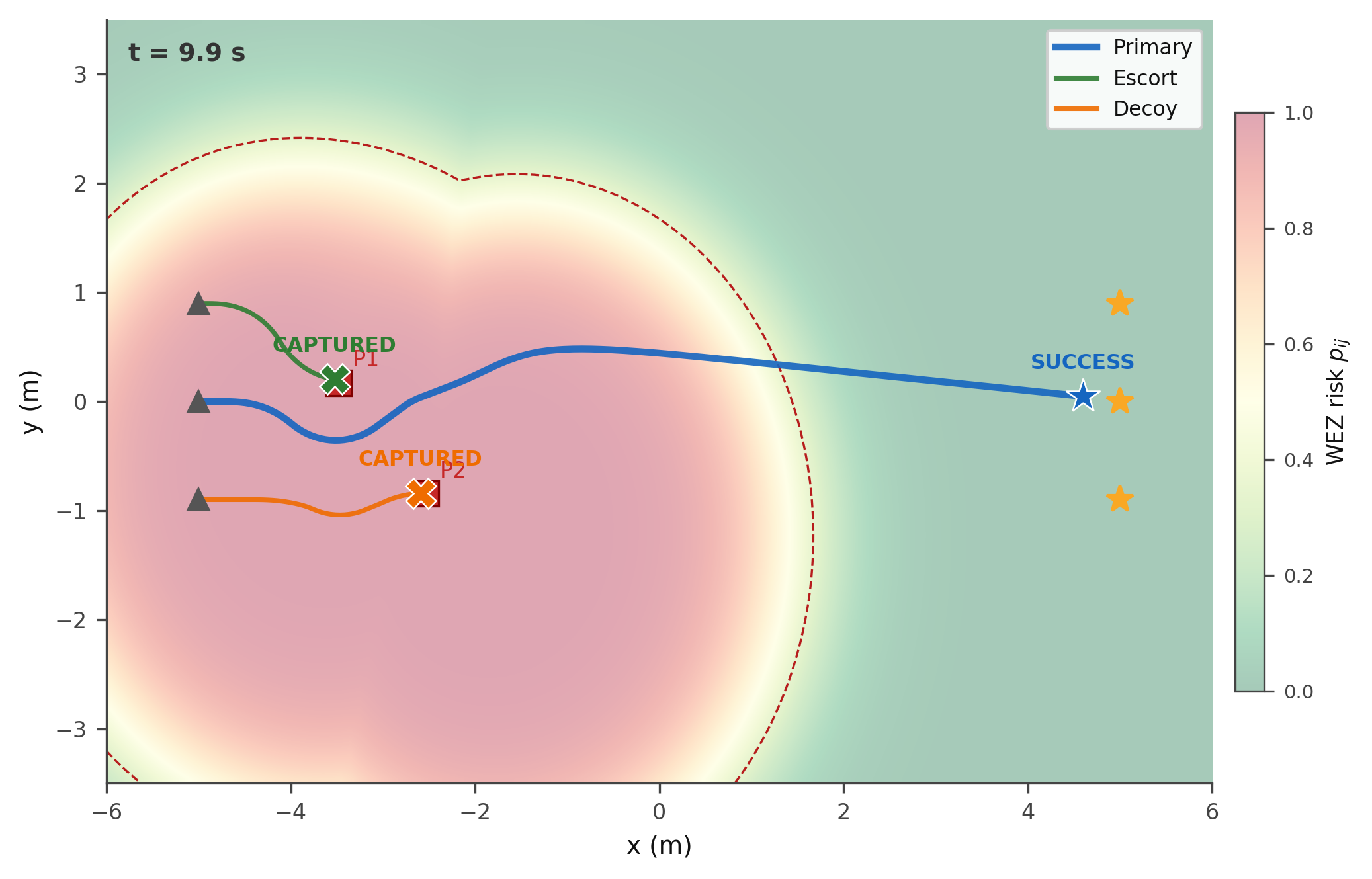}
    \caption{$t\approx11.0$~s: Goal reached. Primary arrives at
             goal; $p\approx0.00$ throughout the final transit leg.}
  \end{subfigure}
  \caption{Keyframe sequence for the reactive-threat experiment.
    Both PurePursuit interceptors (P1, P2) lock onto escort and decoy
    immediately~(a), absorbing them within $2.5$~s~(b), and leaving
    the primary an uncontested corridor~(c--d) --- without any
    explicit inter-vehicle communication.}
  \label{fig:kf_react}
\end{figure}

\Cref{fig:kf_react} illustrates the engagement timeline through four
keyframes.

\begin{figure}[htbp]
  \centering
  \includegraphics[width=0.95\textwidth]{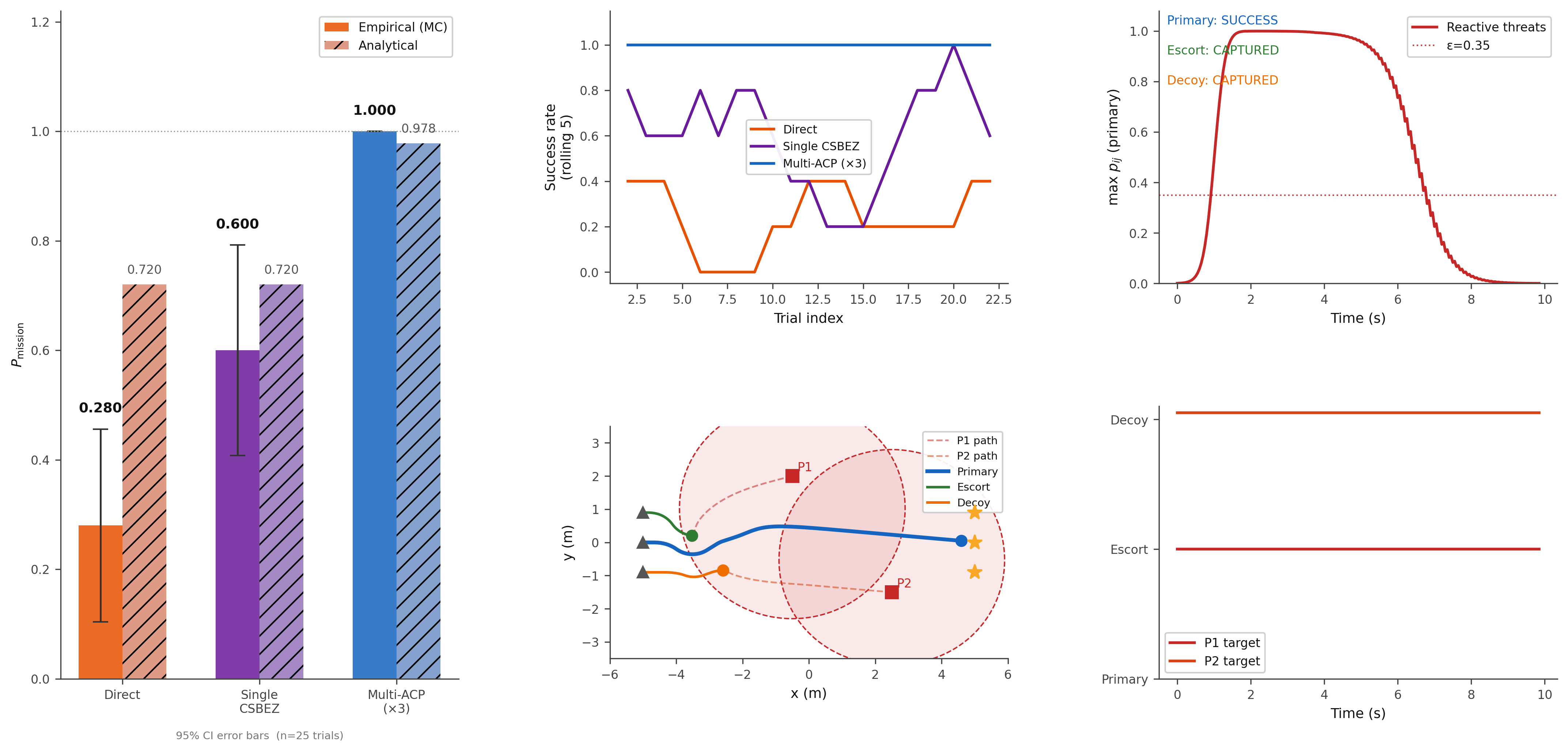}
  \caption{Monte~Carlo and reactive-threat results.
    \textbf{(A)} Empirical $P_\mathrm{mission}$ with 95\% confidence intervals
    (analytical bounds shown hatched): Multi-ACP reaches 1.00 across 100
    trials.
    \textbf{(B)} Rolling 20-trial success rate; Multi-ACP stabilises at 1.00
    from the first window.
    \textbf{(C)} Primary ACP WEZ risk under reactive threats; risk falls to
    near-zero once both threats are occupied by escort and decoy.
    \textbf{(D)} Reactive tactical map: threat paths (dashed) chase escort
    and decoy; primary transits through the cleared corridor.
    \textbf{(E)} Threat target-assignment log: both P1 and P2 assign to
    ACP indices 1--2 (escort, decoy) throughout the engagement and never
    reach index~0 (primary).}
  \label{fig:mc}
\end{figure}

The aggregate Monte~Carlo and reactive-threat results are summarized in
\Cref{fig:mc}.

\section{Discussion and Conclusion}\label{sec:disc}

\subsection{Mechanisms of Improvement}

Three complementary mechanisms account for the gap between single-vehicle
and multi-ACP performance:

\noindent\textbf{Probabilistic redundancy.}
Even on identical routes, $N$ independent attempts raise the team success
probability to $1-(1-p)^N$.
For $p=0.66$ and $N=3$ this alone gives $P=0.96$, before any geometric
advantage is realised.

\noindent\textbf{Threat saturation.}
With $M=2$ interceptors each capable of engaging one target at a time
($C_j=1$), a fleet of $N=3$ can fully occupy all threats while a third
vehicle transits freely.
The escort-capture event in Case~3 is not a failure; it is a
mission-enabling sacrifice consistent with the lower role weight $w=1$.

\noindent\textbf{Route diversity.}
The $\pm0.9$~m lateral offsets separate the three routes beyond the
simultaneous observability radius of any single threat.
This breaks the single-point-of-failure topology of Case~2 and underpins
the reactive-threat result without requiring any explicit decoy command.

\subsection{Limitations and Future Work}

The results presented here establish a strong Phase~I baseline, but several
open directions will be pursued before the January submission.

\noindent\textbf{Probabilistic WEZ and uncertainty propagation.}
The logistic risk model in \Cref{eq:risk,eq:reff} is a convenient surrogate
but is not derived from first principles.
A more principled formulation linearises the CSBEZ scalar field
$z(p_E,\psi_E;\Theta_P)$ around the nominal pursuer parameters
$\bar\Theta_P$ and propagates covariance $\Sigma_\Theta$ through the
gradient as in \Cref{eq:sigma_z}:
\begin{equation}\label{eq:sigma_z}
  \sigma_z^2 = \nabla_\Theta z\big|_{\bar\Theta_P}^\top\,
               \Sigma_\Theta\,
               \nabla_\Theta z\big|_{\bar\Theta_P},
\end{equation}
yielding a Gaussian-backed engagement probability
$p_{ij}=\Phi\!\left(-\bar{z}/\sigma_z\right)$
where $\Phi$ is the standard normal CDF.
This replaces the hand-tuned parameters~$(k,\beta)$ with quantities that
are directly observable (threat range uncertainty, position GPS noise,
speed estimation error), providing a physically interpretable threat model.
Uncertainty propagation will also be applied forward in time via an
unscented transform to obtain predicted WEZ distributions along candidate
waypoints.

\noindent\textbf{Moving intercept targets.}
Current goals are static waypoints.
Replacing them with a moving boat target introduces coupled constraints on
arrival time, path length, and intercept geometry.
The primary ACP must converge to a predicted intercept point that satisfies
both the Dubins reachability constraint and the WEZ safety margin
simultaneously.

\noindent\textbf{Joint pre-planner with blended cost.}
The reactive CSBEZ controller optimises each vehicle independently and
greedily.
A joint pre-planner will minimise the blended cost in
\Cref{eq:joint_cost}:
\begin{equation}\label{eq:joint_cost}
  J = \lambda_1 J_\mathrm{path}
    + \lambda_2 J_\mathrm{WEZ}
    + \lambda_3 J_\mathrm{coord}
    + \lambda_4 J_\mathrm{sync}
    - \lambda_5 P_\mathrm{mission}
\end{equation}
over a candidate waypoint grid shared across all agents, enabling globally
coordinated route deconfliction rather than emergent separation.
Weights $\lambda_i$ will be tuned via sensitivity analysis on the Monte~Carlo
ensemble.

\noindent\textbf{Dynamic role re-assignment.}
Roles (primary, escort, decoy) are fixed at mission start.
An online re-assignment policy triggered by capture events or WEZ-margin
crossings could recover a new primary from the surviving escort, improving
robustness when the fleet is depleted mid-mission.

\noindent\textbf{Heterogeneous threat types.}
The current scenario uses identical pursuers.
Future work will include heterogeneous threats (long-range sensor vs.\
fast interceptor) and will explore how the CSBEZ geometry changes when
the pursuer minimum-turn radius and engagement range are drawn from a
distribution rather than set to nominal values.

\noindent\textbf{Higher-fidelity dynamics and hardware validation.}
The current unicycle model will be replaced with a six-degree-of-freedom
fixed-wing dynamics model to assess whether the CSBEZ reactive controller
remains effective under realistic aerodynamic constraints.
Subsequent steps will address the translation of validated trajectories
into autopilot-compatible waypoint sequences for hardware validation.

\subsection{Conclusion}

This paper has demonstrated that role-differentiated, spatially separated
multi-ACP teams with independent CSBEZ reactive guidance achieve substantially
higher mission success probability than a single WEZ-aware vehicle.
The $+35.8$~percentage-point improvement from $P_\mathrm{mission}=0.72$
to $0.978$ is confirmed both analytically and empirically across 100~Monte
Carlo trials with perturbed threat parameters, and the result extends to
reactive PurePursuit adversaries through the passive geometric deception
mechanism inherent to route separation and role assignment.


\bibliography{threat_aware_autonomy_refs}
\end{document}